\documentclass[12pt]{iopart}

\usepackage{iopams}
\usepackage{setstack}
\usepackage{hyperref}

\def\be{\begin{equation}}
\def\ee{\end{equation}}
\def\ba{\begin{eqnarray}}
\def\ea{\end{eqnarray}}

\begin{document}

\title{Consistency equations in Randall--Sundrum cosmology: a test for braneworld inflation}
\author{Gianluca Calcagni}
\address{Dipartimento di Fisica `M. Melloni', Universit\`{a} di Parma, Parco Area delle Scienze 7/A, I-43100 Parma, Italy}
\ead{calcagni@fis.unipr.it}

\begin{abstract}
In the context of an inflationary Randall--Sundrum Type II
braneworld (RS2) we calculate spectral indices and amplitudes of
cosmological scalar and tensor perturbations, up to second order
in slow-roll parameters. Under very simple assumptions,
extrapolating next-order formulae from first-order calculations in
the case of a de Sitter brane, we see that the degeneracy between
standard and braneworld lowest-order consistency equations is
broken, thus giving different signatures of early-universe
inflationary expansion. Using the latest results from WMAP for estimates
of cosmological observables, it is shown that future data
and missions can in principle discriminate between standard and
braneworld scenarios.
\end{abstract}


\maketitle

\section{Introduction}

The idea that the world we live in has more dimensions than we can see dates back to the 1920s with the works of Kaluza
\cite{kal21} and Klein \cite{kle26}; however, in order not to violate results
coming from gravitational experiments, extra dimensions should be compactified and very tiny (of
order of the Planck scale) and so almost unobservable. During recent years new models have been
proposed (ADD \cite{ADD98}--\cite{ADD99}, RS1 \cite{RSa}, RS2 \cite{RSb}) which require dimensions with
compactification scale close to the limit of modern data ($\lesssim mm$, ADD and RS1 scenarios)
 or even noncompact dimensions (RS2); in these cases, the visible world is confined into a four-dimensional variety
(a brane) embedded in a larger spacetime, thus called braneworld. Besides regaining
classical gravity at low energies, these models have many interesting consequences, such as
 the mass hierarchy problem solution and the concrete possibility to test and bound the theory
by means of astrophysical (supernovae) and cosmological observations (cosmic microwave
background temperature fluctuations, large-scale structures) and accelerators (high-energy
 processes available at future LHCs). However, the problems opened by this new trend of
research are far from being fully explored.

According to the inflationary paradigm, cosmological large-scale structures were originated by
 exponential dilatation of quantum fluctuations up to macroscopic scale, thanks to the
accelerated expansion driven by a scalar field filling up the early universe. The study of
microcosm allows us to investigate macrocosm in some sense, so cosmological observations are
complementary to those with ground-based accelerators. Moreover, they clarify the present
composition and geometrical structure of the universe, as well as those primordial elementary
processes constituting the basis of our visible world.

Our aim is to apply the slow-roll formalism to a single field inflationary model on a brane in order
to compute some observable quantities whose experimental determination from CMB data will be
greatly improved by the WMAP and Planck missions. In particular, we want to compute up to second order
scalar and tensor spectral indices and perturbation amplitudes in analogy with the four-dimensional
standard approach; from these observables we will get the fundamental \emph{consistency equations}
which are typical of inflationary models. According to the lowest-order results (e.g. \cite{HuL}),
 there is no way to discriminate between standard and extra-dimension scenarios,
since the consistency equation is miraculously the same. The hint of our work is that
second-order expressions do break this degeneracy, at least formally. The size of the effect
depends on both physical assumptions and observable quantities, such as the tensor-to-scalar-amplitude ratio. Modern observations turn out to be in the range required for testing the high-energy braneworld cosmology.

This paper is organized as follows: in the rest of this section we
set up the theoretical background; in section 2 we introduce the
slow-roll formalism, making some additional remarks in section 3;
section 4 is devoted to first-order perturbations and in section 5 we extend the analysis to second order by reasonable
guesses; in section 6 observational implications of our results
are considered. Section 7 is devoted to discussion.

\subsection{Theoretical background}
We will consider the RS2 scenario \cite{RSb} in which matter, and in particular the
inflationary scalar field, is confined on a single brane while gravity is free to propagate in
the whole empty bulk. For a complete review of standard inflation, see \cite{lid97}.
The starting point is the continuity equation,
\be
\dot{\rho}+3H (\rho+p) = 0\,,
\ee
and the modified Friedmann equations \cite{CGS}--\cite{BDEL},
\ba
H^2 = \frac{\kappa^2}{6 \lambda} \rho (2 \lambda+\rho)+\frac{\cal E}{a^4}\,,\label{FE1}\\
\frac{\ddot{a}}{a} = -\frac{\kappa^2}{6 \lambda} [\lambda
(\rho+3p)+\rho(2\rho+3p)]-\frac{\cal E}{a^4}\,,\label{FE2}
\ea
where $H$ is the
Hubble parameter restricted to the 3-brane, $\kappa^2 \equiv 8\pi/m^2_4$ includes the effective Planck mass $m_4 \simeq
10^{19}\,{\rm GeV}$, $\lambda$ is the brane tension, $\rho$ is the matter density and ${\cal E}={\rm const.}$ is the
\emph{dark radiation} term which is the time-time component of the
five-dimensional Weyl tensor projected on the brane. Gravity experiments impose the bulk curvature scale to be $\lesssim 1\,{\rm mm}$, that
 is $\lambda^{1/4}>100\,{\rm GeV}$. For a homogeneous scalar field $\phi(t)$ with potential $V$ we have
\ba
\rho &=& \case{1}{2} \dot{\phi}^2 + V(\phi)\,, \label{rho}\\
p &=& \case{1}{2} \dot{\phi}^2 - V(\phi)\,. \label{p}
\ea
In the following we will forget the dark radiation term since during inflation it is strongly
suppressed by the exponential expansion \cite{LMSW,mar01}.

On the brane, the energy-momentum tensor of the scalar field is covariantly conserved,
$\nabla^\nu T_{\mu \nu}=0$, and the field satisfy the same 4D Klein-Gordon equation,
\be \label{eom}
\ddot{\phi} + 3 H \dot{\phi} + V^\prime (\phi) = 0\,.
\ee
A necessary and sufficient condition for inflation to start is $\ddot{a}>0$, which is equal
to a constraint for the equation of state $p=w\rho$,
\be \label{w}
w<-\frac{1}{3}\left(\frac{\lambda+2\rho}{\lambda+\rho}\right)\,,
\ee
which implies
\be \label{ic}
\dot{\phi}^2<V-\frac{\dot{\phi}^2+2V}{8 \lambda} (5 \dot{\phi}^2-2V)\,.
\ee
Finally, we note that
\be \label{dotH}
\dot{H}= - \frac{\kappa^2}{2 \lambda} (\lambda+\rho)(\rho+p)=
- \frac{\kappa^2}{4 \lambda} \dot{\phi}^2 [\dot{\phi}^2+2(\lambda+V)]\,.
\ee


\section{Slow-roll parameters}

The formalism of slow-roll parameters (hereafter denoted by SR) gives good control over the theoretical shape and
amplitude of cosmological perturbations. During the calculation, formulae
containing the smallest power of any of these parameters will be
referred to as `first order SR'. In first order SR it is possible
to drop the second derivative in the equation of motion
(\ref{eom}), $\dot{\phi}\simeq -V'/3H$, the Hubble parameter can
be considered almost constant and the expansion becomes nearly
exponential. Such an approximation has been widely exploited in
braneworld cosmology \cite{HuL,LMW,MWBH} and we will call it extreme slow roll (ESR).

In order to refine previous results it is necessary to push onward to second order SR and
more exactly define the slow-roll parameters. In fact, in the ESR approach one drops subdominant
dynamical terms thus losing information during the system evolution.

We define the following slow-roll parameters:
\ba
\epsilon &\equiv& -\frac{\dot{H}}{H^2} = \frac{3(\lambda+\rho)(\rho+p)}{\rho (\rho+2\lambda)}\,, \label{epsilon}\\
\eta     &\equiv& -\frac{\ddot{\phi}}{H\dot{\phi}} \label{eta}\,,\\
\xi^2    &\equiv& \frac{1}{H^2} \left(\frac{\ddot{\phi}}{\dot{\phi}}\right)^. = \frac{\dddot{\phi}}{H^2\dot{\phi}}-
 \eta^2\,.\label{xi}
\ea
Note that $\epsilon \geq 0$ and, since
\be \label{infl}
\frac{\ddot{a}}{a} = H^2 (1-\epsilon)\,,
\ee
the condition for inflation is precisely $\epsilon < 1$ and inflation ends exactly when
$\epsilon= 1$. Moreover, functional dependence of the parameters on $\phi$ and $H$ is the same
as in the standard case; this is because in both scenarios the definition of inflation is the
same, $\ddot{a}>0$.

The equation of state for the scalar field becomes
\be \label{wepsi}
p= \rho \left[\frac{\rho ({1 \over 3}\epsilon-1)+\lambda (\frac{2}{3}\epsilon-1)}{\rho+\lambda}\right]\,.
\ee


\section{Limits}

In the limit in which the brane tension $\lambda$ is much larger than the scalar field energy density during inflation, the
quadratic term in (\ref{FE1}) is negligible and one recovers the behaviour of the 4D standard cosmological expansion, as one
can see by taking the limit $\lambda/\rho \rightarrow \infty$ in equations (\ref{FE1}),
(\ref{FE2}), (\ref{w}), (\ref{ic}), (\ref{dotH}) and (\ref{wepsi}):
\ba \label{limits}
H^2 \rightarrow \frac{\kappa^2}{3} \rho\,, &\qquad& \frac{\ddot{a}}{a} \rightarrow -\frac{\kappa^2}{6}(\rho+3p)\,,\nonumber\\
w < -\case{1}{3}\,, &\qquad&  \dot{\phi}^2<V\,, \\
\dot{H} \rightarrow -\frac{\kappa^2}{2} \dot{\phi}^2\,, &\qquad& p \rightarrow \left(\frac{2}{3}\,\epsilon-1\right)\rho\,.\nonumber
\ea
In the limit $ \rho \gg \lambda$ there is a deviation from the standard equations which, at least in
 principle, can be measured experimentally.

As has been said before, the slow-roll parameters as functions of $H$ and $\phi$ do not change with respect to the
standard case; however, since the Friedmann equations are different, some relations are no longer
valid (for example, the Hamilton-Jacobi equations) and one must take care of which expression must
be used to define each quantity.

In the ESR limit, the slow-roll parameters become
\ba
\epsilon &\simeq& \frac{m^2_4}{16\pi} \left(\frac{V'}{V}\right)^2 \frac{4 \lambda (V+\lambda)}{(V+2\lambda)^2}\,, \label{epv}\\
\eta &\simeq& \frac{m^2_4}{8\pi} \frac{V''}{V} \frac{2\lambda}{V+2\lambda}\,,\label{etv}
\ea
 as in \cite{MWBH}. In the limit $V \ll \lambda$, equations (\ref{epv}) and (\ref{etv}) reduce to their standard form
 (see, e.g., \cite{LL}).


\section{Cosmological perturbations -- first order SR}

\subsection{Scalar perturbations}

Taking the definitions of \cite{lid97}, the first-order amplitude and spectral index of the scalar perturbations are
\ba
A_S = \frac{1}{5\pi} \left.\frac{H^2}{|\dot{\phi}|}\right|_{k=aH}\,,\\
\ms n_S-1 \equiv \frac{\rmd \ln A_S^2}{\rmd \ln k}\,,
\ea
where $A_S$ is evaluated when the wave number $k$ of the perturbation crosses the horizon during
inflation. Notice that the amplitude has the same $H$ and $\dot{\phi}$ dependence as in four
dimensions; this is due to local conservation of the stress-energy tensor on the brane \cite{MWBH}.
Defining the number of e-foldings as $N(\phi)=\int^{t_e}_{t(\phi)} H(t')
\rmd t'$, where $t_e$ marks  the end of inflation, $k$ can be written as $k(\phi)=H(\phi)a(\phi)=a_e
 H(\phi) \exp[N(\phi)]$, leading to the exact relation
\be
\frac{\rmd \ln k}{\rmd\phi}=\frac{H}{\dot{\phi}}(1-\epsilon)\,.
\label{useful}
\ee
Thus, to first order the spectral index is the standard one,
\be
\label{ns1} n_S-1 = 2 \eta-4\epsilon\,.
\ee

\subsection{Tensor perturbations and the consistency equation}

As regards tensor perturbations, one has \cite{LMW}
\ba
A_T = \frac{2}{5\sqrt{\pi}} \left.\frac{H}{m_4}F\left(\frac{H}{\mu}\right)\right|_{k=aH}\,, \label{A_T1} \\
\ms n_T \equiv \frac{\rmd \ln A_T^2}{\rmd \ln k}\,,
\ea
where $\mu=\sqrt{4\pi\lambda/(3 m_4^2)}$ and 
\[
F(x)=\left[\sqrt{1+x^2}-x^2 \ln \left(1/x+\sqrt{1+1/x^2}\right)\right]^{-1/2}\,.
\] 
In the limit $\rho \ll \lambda$, $F(x) \simeq 1$ and one recovers the standard spectrum:
\ba
A_T = \frac{2}{5\sqrt{\pi}} \left.\frac{H}{m_4}\right|_{k=aH}\,,\\
\ms n_T = -2\epsilon\,.
\ea
The consistency equation is, noting that $A_T^2 / A_S^2 \simeq \epsilon$,
\be \label{ce1}
n_T = -2 \frac{A_T^2}{A_S^2}\,.
\ee
In the limit $\rho \gg \lambda$,  $F(x) \simeq \sqrt{3x/2}$ and
\ba
A_T =  \left.\left(\frac{6 H^3}{25\pi m_4}\right)^{1/2}\left(\frac{3}{4\pi\lambda}\right)^{1/4}\,\right|_{k=aH}\,,\\
\ms n_T = -3\epsilon\,.
\ea
As one can see, the tensor index is different from the previous case and the amplitude is larger
because Kaluza--Klein massive modes contribute in a considerable
way in this energy regime; however, and rather surprisingly, the
consistency equation is the same, equation (\ref{ce1}) (see \cite{HuL}).

The consistency equation reflects the common physical origin of scalar and tensor perturbation;
 it is a typical result from inflation that other theories of structure formation are not able to
 reproduce.

\section{Cosmological perturbations -- second order SR}

Assuming that general relativity is valid up to the characteristic scale of the extra dimension,
 several steps have been made towards a full five-dimensional large-scale perturbation theory
\cite{LMSW,mar01,SMS,VDBL,LCML}. An implementation of perturbation theory in a brane-inflationary context is still at an
 early stage. However, it is possible with some care to use the well
known 4D formalism \cite{bar80}--\cite{MFB}. As we want to make a rough estimate of braneworld effects on observable
anisotropies, we can carry out the calculation with little effort by making a somewhat drastic simplification. In
particular, we will neglect any effects coming from both the total anisotropic stress and the Weyl tensor.
The first assumption reduces the degrees of freedom of gauge invariant scalar perturbations in the
longitudinal (conformal Newtonian) gauge \cite{LMSW}. The second one closes
the system of equations allowing us to study brane physics without nonlocal contributions from the
bulk\footnote[1]{If one chooses a conformally flat background, $\rho_{\cal E}= {\cal E} a^{-4}=0$,
the fluctuation of dark radiation suffers an exponential damping during inflation, $\delta
\rho_{\cal E} \propto a^{-4} \sim \exp (-4Ht)$.}.

This oversimplified background is justified by noting that any improvement of the physics would
only confirm the breaking of degeneracy between standard and braneworld consistency equations,
 which is our main point. We repeat once again that the `softness' of this breaking and its
evidence will strongly depend on the physics, so we cannot say hardly anything \emph{a priori}
about its size in more complicated scenarios. However, as a last remark, it is important to
notice that the effect of this extra-physics (fluctuations in Weyl component and anisotropic stress in a
high-energy regime) will be more enhanced at small scales, since in a conformally and spatially flat background it
only mildly affects the density perturbations and spectrum at large scales, $k \ll aH$ \cite{LCML,GRS,GM}. Since inflationary
dynamics dominates the large-scale perturbation spectrum, in our context we will expect to find
results which are close to the `true' answer coming from a complete computation of the full
Einstein equations with boundary conditions.

As we will see soon, the starting point is to find an exact solution
of the cosmological equations such that the SR parameters are constant,
thus ignoring their time derivatives. One then perturbs the
solution with small variations of the parameters. For example,
from equation (\ref{epsilon}) and (\ref{eta}) one gets
\ba
\dot{\epsilon} &=& 2H\epsilon(\epsilon-\eta)-\frac{12\pi}{\lambda m_4^2} \frac{\dot{\phi}^4}{H}\,, \label{dotepsi}\\
\dot{\eta} &=& H (\epsilon\eta-\xi^2)\,,
\ea
infinitesimal if $\epsilon$, $\eta$, $\xi$ and $\dot{\phi}^4/H$ are small enough. Hawkins and
Lidsey \cite{HaL} have found four braneworld exact models. Here we will consider just two of them.

The first one (model 1) has
\ba
a(T) &=& (T+\sqrt{T^2-1})^p\,,\\
\phi(T) &=& \frac{1}{\gamma} \ln (T+\sqrt{T^2-1})\,,
\ea
where $p>1/3$, $\gamma=\sqrt{4\pi/(p m_4^2)}$ and $T=\sqrt{4\pi\lambda/(3 p^2 m_4^2)}~(t-t_0)$ is the rescaled time, with
$t_0$ being an arbitrary integration constant.
Using the notation $q_T=\rmd q/\rmd T$ for the variable $q$, we have
\be
\tilde{H} \equiv \frac{a_T}{a}= \frac{p}{\sqrt{T^2-1}}\,.
\ee
The slow-roll parameters are
\ba
\epsilon =\eta = \frac{1}{p}\frac{T}{\sqrt{T^2-1}}\,,\\
\ms \xi^2 = \frac{1}{p^2}\frac{T^2+1}{T^2-1}\,.
\ea
In the second model (model 2),
\ba
a(T) = (4T^2-1)^{p/2}\,,\\
\ms \cosh [\gamma\phi(T)] = 2T\,,\\
\ms \tilde{H} = \frac{4pT}{4T^2-1}\,,
\ea
and
\ba
\epsilon = \frac{1}{p}\left(1+\frac{1}{4T^2}\right)\,, \label{epsi2}\\
\ms \eta = \frac{1}{p}\frac{4T^2-1}{4T^2+1}\,, \label{eta2}\\
\ms \xi^2 = \frac{1}{p^2}\left(1+\frac{1}{4T^2}\right)\left(\frac{4T^2-1}{4T^2+1}\right)^2\,.\label{xi2}
\ea
In the limit $T \rightarrow \infty$, these solutions tend to power-law inflation \cite{LM},
with $H=p/t$ and $\epsilon=\eta=\xi=1/p$. This occurs both when the brane
tension is very large and at large times. Actually one can fix the integration constant
 $t_0$ such that $\phi(t=0)=0$, whence $t_0 \sim -m_4/\sqrt{\lambda}$; with this choice,
 one can show that the approximation with constant SR parameters is valid at late times, for instance when
$t \gtrsim 10^3\,t_4$, where $t_4 = m_4^{-1}$ is the Planck time. Thus we have shown that
there exist exact solutions with late time constant SR parameters. Anyway, the reader can
convince oneself that the approximation is good by checking the behaviour of the scale factor $a$ near the origin of time
(model 1: $T\simeq 1$; model 2: $T\simeq \case{1}{2}$). In both cases, again, one obtains a power law
($a\simeq T^p$ and $a\sim t^{p/2}$, respectively) which, combined with the asymptotic behaviour
of $\phi(T)$, generates constant SR parameters.

\subsection{Scalar perturbations}

Let us now see what happens to scalar metric fluctuations. The scalar field is
confined on the brane and the effective action giving the equation of motion (\ref{eom})
reproduces standard four-dimensional cosmology. Scalar perturbations, which are related to scalar
field fluctuations, can thus be treated in the standard way, following step by step the
procedure we now briefly sketch (e.g. \cite{lid97}): (1) write the linearly perturbed metric in terms of gauge-invariant
 scalar quantities; (2) compute the effective action of scalar field fluctuation and the
associated equation of motion; (3) write the perturbation amplitude as a function of an exact
 solution of the equation of motion with constant SR parameters; (4) perturb this solution with
small variations of the parameters.

In the extra horizon limit $k/aH \rightarrow 0$, the scalar amplitude for an inflationary model
with constant SR parameters is \cite{LS}
\[
A_S = \left.\frac{2^{\nu-3/2}}{5\pi} \frac{\Gamma(\nu)}{\Gamma(3/2)}\left(\frac{1}{1-\epsilon}\right)^{1/2-\nu}\frac{H^2}{\dot{\phi}}\right|_{k=aH}\,,
\]
where $\nu$ is a combination of SR parameters; it does not matter which of the exact models is
chosen. Let us now perturb about one of the braneworld
solutions with late time constant $\epsilon$, assuming that $\epsilon$, $\eta$, $\xi$ and
$\dot{\phi}^4/H$ are small quantities. It can be shown that in both the models we are considering
one has $\dot{\phi}^4/H \sim t^{-3}$ at late times. It follows that \cite{SL}
\be
A_S = \left.[1-(2C+1)\epsilon+C\eta]\frac{1}{5\pi}\frac{H^2}{\dot{\phi}}\right|_{k=aH}\,,
\ee
where $C\simeq -0.73$ is a numerical constant. The computation of the spectral index gives
\be
\fl n_S-1 = [-4\epsilon+2\eta-8(C+1)\epsilon^2-2C\xi^2+2(5 C+3)\epsilon\eta]+(2 C+1) \frac{24\pi}{\lambda m_4^2}\frac{\dot{\phi}^4}{H^2}[1+O(\epsilon, \eta)]\,.
\ee
The first additive term on the right-hand side is the standard expression for the spectral index, while the second
term comes from the embedding of inflationary cosmology in a five-dimensional bulk. For $\rho
\ll \lambda$ one has, using (\ref{limits}),
\be
\frac{\dot{\phi}^4}{\lambda H^2} \sim \frac{\rho}{\lambda}\epsilon^2 \label{new1}
\ee
and the extra term can be neglected, thus finding the standard result
\be
n_S-1 = -4\epsilon+2\eta-8(C+1)\epsilon^2-2C\xi^2+2(5 C+3)\epsilon\eta\,. \label{n_Sstandard}
\ee
For $\rho \gg \lambda$, from equations (\ref{rho}), (\ref{p}), (\ref{dotH}), (\ref{epsilon}), (\ref{wepsi}) and
(\ref{dotepsi}) we get
\ba
\frac{12\pi}{\lambda m_4^2}\frac{\dot{\phi}^4}{H^2} \simeq \epsilon^2 \label{new2}\,,\\
\ms\dot{\epsilon} = H \epsilon (\epsilon-2\eta)\,, \label{dotepsinew}
\ea
whence
\be
n_S-1= -4\epsilon+2\eta-2(2 C+3)\epsilon^2-2C\xi^2+2(5 C+3)\epsilon\eta\,. \label{n_Snew}
\ee
The only difference with respect to (\ref{n_Sstandard}) is the numerical coefficient in front
of $\epsilon^2$, $-8(C+1) \simeq -2.16$ in the first case and $-2(2 C+3) \simeq -3.08$ in the second case.

\subsection{Tensor perturbations and consistency equations}

Tensor perturbations are more difficult to deal with than scalar
ones since gravity is free to propagate in the whole five-dimensional
spacetime. In general, the graviton zero-mode, localized on the brane,
interacts with Kaluza--Klein massive modes generating an infinite
tower of coupled differential equations. In some limits the
analysis simplifies and the zero-mode decouples; for instance, for
a de Sitter brane ($\rho+p=0$) the problem can be resolved and one
gets equation (\ref{A_T1}) \cite{LMW}.

However, some considerations can give semi-quantitative hints
about the shape of the perturbation amplitude. In fact, we know that
at the lowest order in SR parameters the amplitude is given by
equation (\ref{A_T1}), while in the low-energy limit we must
obtain the standard 4D amplitude which is known to the next-to-lowest SR order \cite{SL}:
\[
A_T = \left.\frac{2}{5\sqrt{\pi}}[1-(C+1)\epsilon]\frac{H}{m_4}\right|_{k=aH}\,.
\]
Our ansatz for tensor amplitude to the second order in SR parameters is then
\be \label{A_Tnew2}
A_T = \left.\frac{2}{5\sqrt{\pi}}[1-(C+1)\epsilon]\frac{H}{m_4}F\left(\frac{H}{\mu}\right)\right|_{k=aH}\,.
\ee
The calculation of spectral index and consistency equations is a simple task in the low ($\rho
\ll \lambda$) and high ($\rho \gg \lambda$) energy limits, labelled by the superscripts {\it
(l)} and {\it (h)}, respectively. In the first case one gets the standard result \cite{SL,CKLL}:
\ba
n_T^{(l)} &=& -\epsilon [2+2(2 C+3) \epsilon - 4 (C+1) \eta] \label{n_Told2} \\
          &=& -2 \frac{A_T^2}{A_S^2}\left[1-\frac{A_T^2}{A_S^2}+(1-n_S)\right] \label{ecnTold}\,.
\ea Since to second order there appear only quantities present in
first-order expressions, it is reasonable to consider equation
(\ref{ecnTold}) not as an extension of (\ref{ce1}) but as \emph{the} consistency equation.

For $\rho \gg \lambda$ one gets
\be
n_T^{(h)} = -\epsilon [3+(2
C+5) \epsilon - 4 (C+1) \eta] \label{n_Tnew2}\,.
\ee
Using the approximate relation
\be
\epsilon \simeq \frac{2}{3}\frac{A_T^2}{A_S^2}[1-2C(\epsilon-\eta)]
\ee
and the first-order equation (\ref{ns1}), we obtain the main result of this work
\ba
n_T^{(h)} &=& -\frac{2}{3} \frac{A_T^2}{A_S^2}\left[3-2\frac{A_T^2}{A_S^2}+(2-C)(1-n_S)\right] \label{ec2}\\
          &=& -2 \frac{A_T^2}{A_S^2}\left[1-0.67\frac{A_T^2}{A_S^2}+0.91(1-n_S)\right]\,.
\ea
The first-order degeneracy between the two regimes is thus (softly) broken. Note that to first
order $n_S^{(h)}=n_S^{(l)}=n_S$.

We can get a second consistency equation by computing $\alpha_T \equiv \rmd n_T/\rmd \ln k$ starting from equations
(\ref{n_Told2}), (\ref{n_Tnew2}), (\ref{dotepsi}), (\ref{dotepsinew}) and (\ref{useful}). For $\rho \ll \lambda$, we
have $\alpha_T \simeq 4 \epsilon (\eta-\epsilon)$, whence, using (\ref{ns1}) and the first-order relation $A_T^2 / A_S^2
\simeq \epsilon$ \cite{KT},
\be
\label{eck} \alpha_T = 2\frac{A_T^2}{A_S^2} \left[2 \frac{A_T^2}{A_S^2}+(n_S-1)\right]\,.
\ee
When $\rho \gg \lambda$, (\ref{dotepsinew}) gives $\alpha_T
\simeq 3 \epsilon (2\eta-\epsilon)$ and so, since $A_T^2 / A_S^2
\simeq 3\epsilon/2$, one finds (\ref{eck}) again. In the light of
the results of the previous section, this coincidence is not very
striking, having used only first-order expressions for the
computation.

A final consistency equation comes from the running of the scalar index, $\alpha_S \equiv
\rmd n_S/\rmd \ln k$. In the low-energy limit one finds $\alpha_S \simeq 2(-4 \epsilon^2+
5\epsilon\eta-\xi^2)$, while in the limit $\rho \gg \lambda$ we have $\alpha_S \simeq 2(-2
\epsilon^2+ 5\epsilon\eta-\xi^2)$. These are expressions quadratic in the SR parameters, that can be
written in terms of observables thanks to the standard assumption $\xi \ll {\rm max}\{\epsilon,
 |\eta|\}$, which permits us to neglect the last term, thus obtaining
\ba
\alpha_S^{(l)} &=& \vphantom{\frac{2}{3}}\,\,\,\,\frac{A_T^2}{A_S^2} \left[12 \frac{A_T^2}{A_S^2}+5 (n_S-1)\right]\,, \label{run1}\\
\alpha_S^{(h)} &=& \frac{2}{3} \frac{A_T^2}{A_S^2} \left[\frac{32}{3} \frac{A_T^2}{A_S^2}+5 (n_S-1)\right]\,.\label{run2}
\ea
For a recent study of the running of the spectral index, see \cite{LT}.

\section{Observational consequences}

Let us now see how much the low and high energy scenarios are different by comparing the consistency equations at the
two energy scales, equation (\ref{ecnTold}), (\ref{ec2}), (\ref{eck})--(\ref{run2}):
\ba
n_T^{(h)}-n_T^{(l)}           = -\frac{2}{3} \frac{A_T^2}{A_S^2}\left[\frac{A_T^2}{A_S^2}+(1+C)(n_S-1)\right]\,, \label{sper1}\\
\alpha_T^{(h)}-\alpha_T^{(l)} = 0\,, \label{sper3}\\
\alpha_S^{(h)}-\alpha_S^{(l)} = -\frac{1}{3}
\frac{A_T^2}{A_S^2}\left[\frac{44}{3}\frac{A_T^2}{A_S^2}+5(n_S-1)\right]\,.\label{sper2}
\ea
The last equation remains valid if we relax the condition $\xi \ll {\rm max}\{\epsilon, |\eta|\}$
 by assuming instead the parameter $\xi$ to be almost constant in the energy scale, $\xi^{(h)}
\simeq \xi^{(l)}$. Note that the running of the tensor index does not give observational evidence of the braneworld.

The first-year data of the {\it Wilkinson Microwave Anisotropy Probe} (WMAP) \cite{ben03}--\cite{spe03}
 confirm the standard scenario of a flat, adiabatic universe with Gaussian, almost scale-invariant anisotropies.
 Bennett \etal \cite{ben03} put a bound on the amplitudes' ratio, $r=16 A_T^2/A_S^2<0.90$;
here we will take the upper limit $A_T^2/A_S^2=0.06$ and the best
fit for the scalar index of \cite{spe03}, $n_S=0.93$. Substituting
these values for the observables in equation (\ref{sper2}), we
have
\be
\left|\alpha_S^{(h)}-\alpha_S^{(l)}\right| \simeq 0.01\,,
\ee
comparable both with the error in the estimate of Bridle \etal
\cite{bri03}, $\alpha_S=-0.04 \pm 0.03$, coming from the
combination of WMAP and 2dFGRS ({\it 2 Degree Field Galaxy
Redshift Survey}) data, and with the uncertainty estimate of the
Planck mission \cite{CGL}. However, we will get a much smaller
effect if the contribution of tensor perturbations is
significantly suppressed.

Data analysis carried out for WMAP makes use of the consistency equation (\ref{ce1}) in order to fix the tensor index and its
running in the space of parameters, so a direct confrontation between these quantities and an experimental result with an
associate error is not possible. However, taking the above estimates and putting $n_T \simeq -0.1$, equation (\ref{sper1})
 gives $\big|(n_T^{(h)}-n_T^{(l)})/n_T\big| \simeq 0.02$; testing this
effect would require an experimental uncertainty less than 1\% for the
tensor index, a very difficult goal to hit for the missions of this generation.

As a last comment we note that the comparison of the observable
quantities $n_T$, $\alpha_T$ and $\alpha_S$ must be done with the
consistency equations, equations (\ref{sper1})--(\ref{sper2}) and not through the SR parameters expressions, equations
(\ref{n_Told2}), (\ref{n_Tnew2}) and following. This is both
because we are dealing with independent expressions and because
there is an evident ambiguity in relations between SR parameters
and observables when considering energy-scale finite differences
of the quantities of interest. For example, one would have
$\alpha_S^{(h)}-\alpha_S^{(l)}=4\epsilon^2$, where $\epsilon$
would be arbitrarily evaluated along the energy scale.

Further braneworld effects on the CMB spectrum and confrontation with WMAP data are studied e.g. in \cite{BFM}.

\section{Discussion}

Within the noncompact braneworld Randall--Sundrum model, we have
computed the main theoretical quantities describing the cosmological
perturbations generated from a brane-confined inflationary period.
In order to find the braneworld signatures in cosmological
observations we have carried out a semi-quantitative investigation
starting from previous de Sitter results instead of doing actual
5D calculations, which are very complicated due to
 the presence of the full tower of gravitational modes (that is,
 the brane is not an effectively closed 4D system). In particular, we have written spectral indices and consistency
equations in the low-energy and high-energy limits using slow-roll
parameters, to lowest and next-to-lowest order.

To first order in the SR parameters, the braneworld consistency equation
is the same as the standard one, equation (\ref{ce1}), and so it
is not possible to distinguish between the two scenarios. To
second order, however, the consistency equations are formally
different in the two regimes, thus permitting us, in principle, to find
observational clues in favour of the braneworld model. Even if the present
data do not allow such a discrimination, new data coming from WMAP
and the future Planck satellite could throw more light on the
problem.

Finally, there are some generalizations that can give new details of the cosmological perturbation
structures. In particular, one can of course relax the condition of vanishing Weyl
 or energy-momentum tensor in the bulk \cite{LMSW,LCML},
allowing the presence of matter outside the brane too. For the case of five-dimensional
inflation, see \cite{HS01}--\cite{MHS}.

\ack

The author thanks Z Berezhiani for his kind hospitality at Gran Sasso National Laboratories,
L Griguolo, S Matarrese and M Pietroni for useful discussions and A Trevisiol for moral
support. I am also grateful to A R Liddle who draw my attention to his recent paper with E Ram\'{i}rez \cite{RL}.

\end{document}